\documentclass[twocolumn,prb,showpacs,superscriptaddress,preprintnumbers,amsmath,amssymb,floatfix]{revtex4}
\usepackage{graphicx}
\usepackage{dcolumn}
\usepackage{bm}
\usepackage{color}
\usepackage{ragged2e}
\usepackage{adjustbox}
\usepackage{color}

\begin{document}

\title{Probing the $J_{eff}=0$ ground state and the Van Vleck paramagnetism of the Ir$^{5+}$
ions in the layered Sr$_2$Co$_{0.5}$Ir$_{0.5}$O$_4$}
\author{S.~Agrestini}
 \affiliation{Max Planck Institute for Chemical Physics of Solids,
     N\"othnitzerstr. 40, 01187 Dresden, Germany}
 \affiliation{ALBA Synchrotron Light Source, E-08290 Cerdanyola del Vall\`{e}s, Barcelona, Spain}
\author{C.-Y.~Kuo}
 \altaffiliation[Present address: ]{National Synchrotron Radiation Research Center, 101 Hsin-Ann Road, Hsinchu 30076, Taiwan}
 \affiliation{Max Planck Institute for Chemical Physics of Solids,
     N\"othnitzerstr. 40, 01187 Dresden, Germany}
\author{K.~Chen}
 \altaffiliation[Present address: ]{Synchrotron SOLEIL, L'Orme des Merisiers, Saint-Aubin, 91192 Gif-sur-Yvette, France}
 \affiliation{Institute of Physics II, University of Cologne, Z\"ulpicher Stra{\ss}e 77, 50937 Cologne, Germany}
\author{Y.~Utsumi}
 \altaffiliation[Present address: ]{Synchrotron SOLEIL, L'Orme des Merisiers, Saint-Aubin, 91192 Gif-sur-Yvette, France}
 \affiliation{Max Planck Institute for Chemical Physics of Solids,
     N\"othnitzerstr. 40, 01187 Dresden, Germany}
\author{D.~Mikhailova}
 \affiliation{Max Planck Institute for Chemical Physics of Solids,
     N\"othnitzerstr. 40, 01187 Dresden, Germany}
 \affiliation{Leibniz Institute for Solid State and Materials Research (IFW) Dresden e.V., Helmholtzstraße 20, D-01069 Dresden, Germany}
\author{A.~Rogalev}
 \affiliation{ESRF-The European Synchrotron, 71 Avenue des Martyrs, 38000 Grenoble, France}
\author{F.~Wilhelm}
 \affiliation{ESRF-The European Synchrotron, 71 Avenue des Martyrs, 38000 Grenoble, France}
\author{T. F\"orster}
 \affiliation{Hochfeld-Magnetlabor Dresden (HLD-EMFL), Helmholtz-Zentrum Dresden-Rossendorf, 01314 Dresden, Germany}
\author{A.~Matsumoto}
 \affiliation{Department of Physics and Department of Advanced Materials, University of Tokyo, 7-3-1 Hongo, Tokyo 113-0033, Japan}
\author{T.~Takayama}
 \affiliation{Department of Physics and Department of Advanced Materials, University of Tokyo, 7-3-1 Hongo, Tokyo 113-0033, Japan}
 \affiliation{Max Planck Institute for Solid State Research, Heisenbergstrasse 1, 70569 Stuttgart, Germany}
\author{H.~Takagi}
 \affiliation{Department of Physics and Department of Advanced Materials, University of Tokyo, 7-3-1 Hongo, Tokyo 113-0033, Japan}
 \affiliation{Max Planck Institute for Solid State Research, Heisenbergstrasse 1, 70569 Stuttgart, Germany}
 \affiliation{Institute for Functional Matter and Quantum Technologies, University of Stuttgart, Pfaffenwaldring 57, 70569 Stuttgart, Germany}
\author{M.~W.~Haverkort}
 \affiliation{Max Planck Institute for Chemical Physics of Solids, N\"othnitzerstr. 40, 01187 Dresden, Germany}
 \affiliation{Institute for theoretical physics, Heidelberg University, Philosophenweg 19, 69120 Heidelberg, Germany}
\author{Z.~Hu}
  \affiliation{Max Planck Institute for Chemical Physics of Solids,
     N\"othnitzerstr. 40, 01187 Dresden, Germany}
\author{L.~H.~Tjeng}
  \affiliation{Max Planck Institute for Chemical Physics of Solids,
     N\"othnitzerstr. 40, 01187 Dresden, Germany}

\date{\today}
\begin{abstract}
We report a combined experimental and theoretical x-ray magnetic circular dichroism (XMCD)
spectroscopy study at the Ir-$L_{2,3}$ edges on the Ir$^{5+}$ ions of the layered hybrid solid state oxide Sr$_2$Co$_{0.5}$Ir$_{0.5}$O$_4$ with the K$_2$NiF$_4$ structure. From theoretical simulation of the experimental Ir-$L_{2,3}$ XMCD spectrum, we found a deviation from a pure $J_{eff}=0$ ground state with an anisotropic orbital-to-spin moment ratio ($L_x/2S_x$ = 0.43 and $L_z/2S_z$ = 0.78). This deviation is mainly due to multiplet interactions being not small compared to the cubic crystal field and due to the presence of a large tetragonal crystal field associated with the crystal structure. Nevertheless, our calculations show that the energy gap between the singlet ground state and the triplet excited state is still large and that  the magnetic properties of the Ir$^{5+}$ ions can be well described in terms of singlet Van Vleck paramagnetism.

\end{abstract}


\maketitle

\section{Introduction }

The class of iridium based oxides has attracted tremendous attention in recent years.
The presence of strong spin-orbit coupling (SOC) in the 5d shell and associated entanglement
of the spin and orbital degrees of freedom may lead to unexpected exotic electronic states.
In the present work, we focus on the layered Sr$_2$Co$_{0.5}$Ir$_{0.5}$O$_4$, a material
which we were able to synthesize as a single phase and stoichiometric without oxygen deficiency.
The two parent compounds Sr$_2$IrO$_4$ and Sr$_2$CoO$_4$ have very different physical
properties, despite having the same crystal structure. Sr$_2$IrO$_4$ is a canted antiferromagnet
with $T_N$ = 240 K, where the strong SOC leads to an insulating state \cite{Kim.2008}.
The pseudospin $J_{eff}=1/2$ state has been proposed as the ground state of the Ir$^{4+}$ ions
in Sr$_2$IrO$_4$ \cite{Kim.2008}, with the magnetic interactions described by a Heisenberg model
\cite{Jackeli.2009} akin to the spin-1/2 Hamiltonian used to represent the magnetic dynamics in
La$_2$CuO$_4$. Hence, electron doped Sr$_2$IrO$_4$ has raised a huge interest as possible
analogue of the hole doped cuprate high-temperature superconductors \cite{Wang.2011}.
Sr$_2$CoO$_4$, instead, is a metallic ferromagnet ( $T_c$ = 250 K) with the Co$^{4+}$ ions in
the $S=1$ spin state \cite{Matsuno.2004}. Due to the very different electronic and magnetic
properties of the two end compounds, Sr$_2$IrO$_4$ and Sr$_2$CoO$_4$, the solid state solution
Sr$_2$Co$_{x}$Ir$_{1-x}$O$_4$ is expected to exhibit interesting physics.

In an early study of the Sr$_2$Co$_{x}$Ir$_{1-x}$O$_4$ system, where the authors were able
to replace only 30\% of Ir ions by Co ions, the observed increase of the effective magnetic
moment was interpreted in terms of Ir$^{4+}$/Co$^{4+}$ valence states with Co ions in the
intermediate spin state similar to Sr$_2$CoO$_4$ \cite{Gatimu.2012}. However, a theoretical work
\cite{Ou.2014} proposed that the introduced Co ions in the Sr$_2$IrO$_4$ matrix would prefer
to have a lower valence state (3+) than the 4+ high valence and hence induce a charge state
change of the Ir ions. Indeed, following our recent successful synthesis of
Sr$_2$Co$_{0.5}$Ir$_{0.5}$O$_4$, we were able to find experimental evidence in support of the
Ir$^{5+}$/Co$^{3+}$ scenario \cite{Mikhailova.2017,Agrestini.2017}. Such a scenario is very
interesting, because the magnetic ground state of Ir$^{5+}$ has been recently subject of debate
\cite{Khaliullin.2013,Meetei.2015,Cao.2014,Dey.2016,Corredor.2017}. In the limit of strong SOC
and large on-site Coulomb energy $U$, ions with $5d^4$-$t_{2g}^4$ configuration are expected
to be in the $J_{eff}=0$ ground state. However, the large SOC and $U$ limit has been questioned
by theoretical studies, which proposed that strong inter-site hopping may lead to large enough
superexchange interactions resulting in a magnetic condensation of Van-Vleck excitons \cite{Khaliullin.2013}
and novel magnetic states \cite{Meetei.2015}. Whether the recently reported low-temperature
antiferromagnetic order in Sr$_2$YIrO$_6$ \cite{Cao.2014} is related to this novel magnetism is
a topic under current debate \cite{Dey.2016,Corredor.2017}.

\section{Experimental }

Synthesis of the layered Sr$_{2}$Co$_{0.5}$Ir$_{0.5}$O$_{4}$ was carried out from stoichiometric
powder mixtures of home-made Co$_3$O$_4$ with IrO$_2$ (Umicore) and SrCO$_3$ (Alfa Aesar, 99.99\%)
at 1200~$^{\circ}$C in air for 80~h. Co$_3$O$_4$ was obtained by thermal decomposition of
Co(NO$_3$)$_2$*6H$_2$O at 700~$^{\circ}$C in an oxygen flow. In order to obtain fully oxidized
Sr$_2$Co$_{0.5}$Ir$_{0.5}$O$_4$ samples for spectroscopic studies, post-annealing in steel autoclaves
at 400~$^{\circ}$C and 5000~bar O$_2$ pressure was performed for five days. The phase analysis and
the determination of the unit cell parameters were performed using x-ray powder diffraction~\cite{Mikhailova.2017}. Transition metal cations Co and Ir are in edge-sharing oxygen octahedra that are elongated along the c-axis. Single crystals of Sr$_2$IrO$_4$ were grown by the flux method.

The XMCD spectra at the Ir-$L_{2,3}$-edges were measured at the beamline ID12 \cite{Rogalev.2015} of ESRF in Grenoble
(France) with a degree of circular polarization of about 97\%. Spectra were recorded using the bulk sensitive total fluorescence yield detection mode. The XMCD signal was measured in a magnetic field of 17~Tesla with the sample kept at a temperature of 2~K. The Ir-$L_{2,3}$ x-ray absorption spectra for right and left circularly polarized light were corrected for self-absorption effects. The Ir $L_3/L_2$ edge-jump intensity ratio $I(L_3)/I(L_2)$ was then normalized to 2.22 \cite{Henke.1993}. This takes into account the difference in the radial matrix elements of the $2p_{1/2}$-to-$5d(L_2)$ and $2p_{3/2}$-to-$5d(L_3)$ transitions. Magnetization measurements in pulsed fields up to 58~T where made using a pair of compensated pickup coils. The pulsed field data was than scaled with low field data obtained by SQUID MPMS magnetometer.

\section{Results}

\begin{figure}[t]\centering
\includegraphics[width=0.9\linewidth]{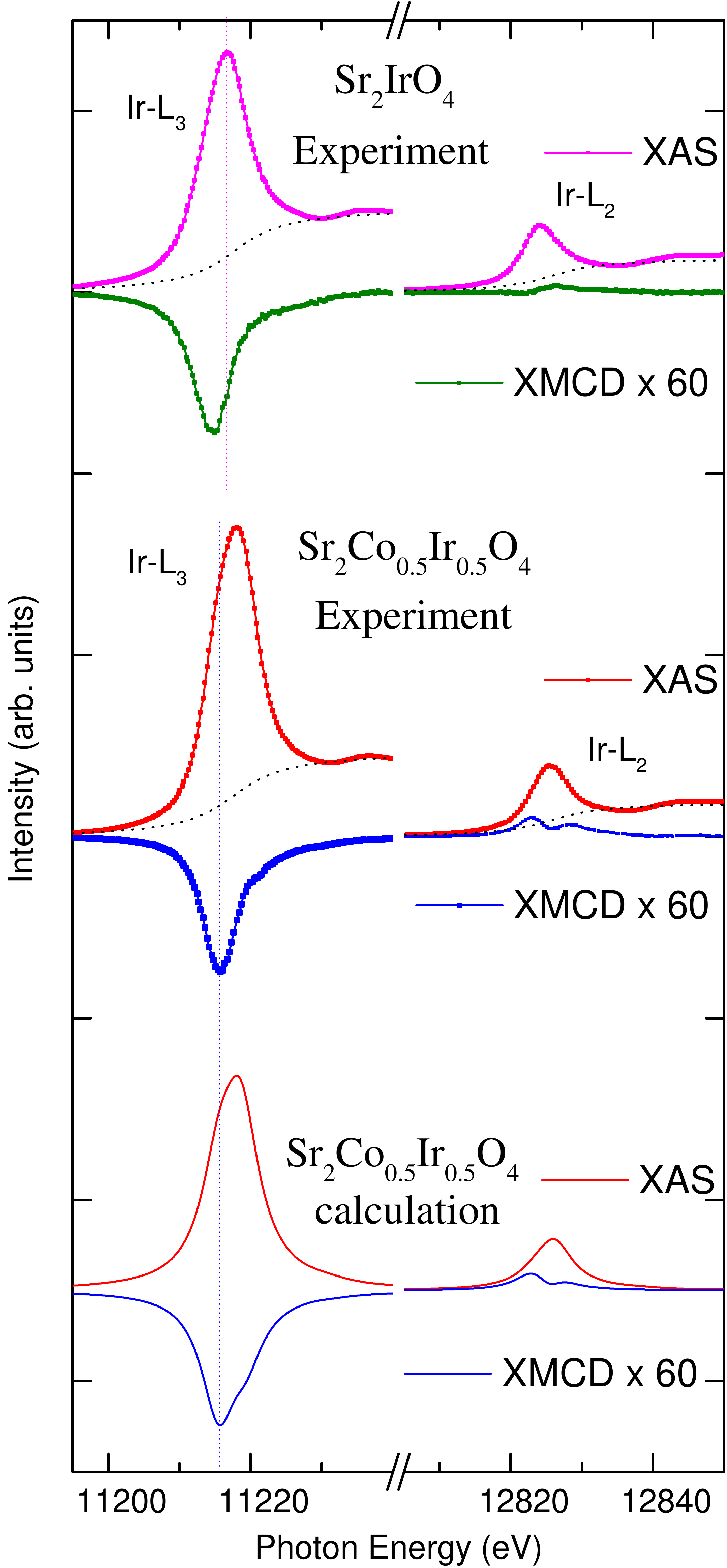}
\caption{Experimental Ir-$L_{2,3}$ XAS and XMCD spectra of Sr$_2$Co$_{0.5}$Ir$_{0.5}$O$_4$,
red and blue circles, respectively, and of Sr$_2$IrO$_4$, magenta and green circles, respectively.
The spectra were measured at $T$ = 2~K and $B_{app}$ = 17~T. The vertical dotted lines illustrate
the energy shift between the spectra of the two samples. The black dotted curves represent the edge jumps. Bottom: calculated Ir-$L_{2,3}$ XAS (red line) and XMCD (blue line) of Sr$_2$Co$_{0.5}$Ir$_{0.5}$O$_4$.}\vspace{-0.2cm}
\label{fig.1}
\end{figure}

As a first step in our investigation of Sr$_2$Co$_{0.5}$Ir$_{0.5}$O$_4$, one needs to make sure
that the Ir magnetism probed by the XMCD technique originates from the Ir$^{5+}$ ions and not
from Ir$^{4+}$ ions impurities. For this purpose we have performed XMCD measurements also on
pure Sr$_2$IrO$_4$, as reference for Ir$^{4+}$ ions sitting on the same local environment as in
the investigated compound. 

In Fig.~\ref{fig.1} we report the results of the Ir-$L_{2,3}$ X-ray absorption spectroscopy experiments carried out on a polycrystalline pellet of Sr$_2$Co$_{0.5}$Ir$_{0.5}$O$_4$ and on a single crystal of Sr$_2$IrO$_4$ at $T$ = 2~K and magnetic field $B_{app}$ = 17~T. The X-ray absorption spectra were taken using circular polarized light with the photon helicity being either parallel or antiparallel with respect to the applied magnetic field. The difference spectrum, called XMCD, and the sum spectrum, called XAS, of Sr$_2$Co$_{0.5}$Ir$_{0.5}$O$_4$ ( Sr$_2$IrO$_4$ ) are shown in Fig.~\ref{fig.1} as blue and red (magenta and green) curves, respectively. In the case of Sr$_2$IrO$_4$ the spectra were collected
in both normal and grazing incidence with the magnetic field forming an angle of $90^{\circ}$
($B\bot ab$) and $20^{\circ}$ ($B//ab$), respectively, with the $ab$ plane. The XMCD of
Sr$_2$IrO$_4$ measured for $B\bot ab$ is very tiny, 20 times smaller in size than the XMCD
measured for $B//ab$, in agreement with the large magnetic anisotropy revealed by magnetization
measurements\cite{Takayama.2016}. In Fig.~\ref{fig.1} we show only the Sr$_2$IrO$_4$ data
collected in grazing incidence. The XMCD spectrum of Sr$_2$IrO$_4$ is in very good agreement
with the data published in literature\cite{Haskel.2012}.

It is important to notice that at both the Ir-$L_{3}$ and the Ir $L_{2}$ edges the XAS of
Sr$_2$Co$_{0.5}$Ir$_{0.5}$O$_4$ lies about 1.3~eV higher in energy than the XAS of Sr$_2$IrO$_4$,
as illustrated by the vertical dotted lines in Fig.~\ref{fig.1}. It is well known that x-ray absorption
spectra at the transition-metal $L_{2,3}$ edges are highly sensitive to the valence state. An increase of the valence of
the metal ion by one results in a shift of the $L_{2,3}$ XAS spectra to higher energies by 1 eV or more\cite{Chen.1990,Choy.1995,Mitra.2003,Burnus.2006,Burnus.2008,Baroudi.2014}. This shift is due to a final state effect in the x-ray absorption process. The energy difference between a $5d^n$ ($5d^5$ for Ir$^{4+}$) and a $5d^{n-1}$ ($5d^4$ for Ir$^{5+}$) configuration is $\Delta$$E = E(2p^6 5d^{n-1} \rightarrow 2p^5 5d^n) - E(2p^6 5d^n \rightarrow 2p^5 5d^{n+1}) = U_{pd} - U_{dd} = $ 1-2 eV, where $U_{dd}$ is the Coulomb repulsion energy between two 5$d$ electrons and $U_{pd}$ the one between a 5$d$ electron and the 2$p$ core hole. The difference in energy position of the XAS, hence, confirms that the 50\% replacement of the Iridium ions with Cobalt ions in the Sr$_2$IrO$_4$ matrix induces an increase of the valence of the remaining Iridium ions from 4+ to 5+. 

The XMCD signal of Sr$_2$Co$_{0.5}$Ir$_{0.5}$O$_4$ at the Ir-$L_{3}$ exhibits a similar lineshape as that of
Sr$_2$IrO$_4$. One might then wonder whether the XMCD signal of Sr$_2$Co$_{0.5}$Ir$_{0.5}$O$_4$
is due to Ir$^{4+}$ impurities considering that the Ir$^{5+}$ ions should be Van-Vleck ions in the
strong spin-orbit coupling limit. However, by looking carefully at the XMCD spectra one can notice that the energy position is different: the Ir-$L_{3}$ XMCD peak of Sr$_2$Co$_{0.5}$Ir$_{0.5}$O$_4$ occurs at about 1.1 eV higher energies than that of Sr$_2$IrO$_4$. Furthermore, the XMCD spectra of the two samples at the Ir-$L_{2}$ edge have completely different lineshape: Sr$_2$Co$_{0.5}$Ir$_{0.5}$O$_4$ shows a double peak feature with positive intensity, while Sr$_2$IrO$_4$ exhibits only one peak with positive intensity on the high energy side of the Ir-$L_{2}$, and a slight negative intensity on the low energy side. These differences in energy position at the Ir-$L_{3}$ edge and in spectral lineshape at the Ir-$L_{2}$ edge demonstrate that the XMCD signal of Sr$_2$Co$_{0.5}$Ir$_{0.5}$O$_4$ cannot be due to the presence of Ir$^{4+}$ impurities, but is related to the field induced magnetism of the Ir$^{5+}$ ions. We also would like to note that our Sr$_2$Co$_{0.5}$Ir$_{0.5}$O$_4$ XMCD spectrum has different details in the line shape in comparison to the ones reported for Sr$_2$Fe$_{0.5}$Ir$_{0.5}$O$_4$ and Sr$_2$In$_{0.5}$Ir$_{0.5}$O$_4$ \cite{Laguna.2015}.

The large difference in intensity of the dichroic signal between the Ir $L_3$ and $L_2$ edges shown in
Fig.~\ref{fig.1} indicates clearly that the Ir ions have a significant unquenched orbital moment
\cite{Thole.1992}.  In order to extract directly from the spectrum the ratio of orbital and spin moments
we have used the sum rules for XMCD developed by Thole et al. \cite{Thole.1992} and Carra et al.
\cite{Carra.1993}. The sum rules can be summarized as:

\begin{equation}\label{eq:ratio}
\frac{L_z}{2S_z+7T_z}=\frac{2}{3}\cdot \frac{\int_{L_{2,3}}(\sigma^+-\sigma^-)dE}{\int_{L_{3}}(\sigma^+-\sigma^-)dE-2\int_{L_{2}}(\sigma^+-\sigma^-)dE}.
\end{equation}

where $S_z$ and $L_z$ are the spin and orbital contributions to the local magnetic moment, respectively, and $T_z$ is the intra-atomic magnetic dipole moment. Advantage of this sum rule is that it does not require a saturation of the magnetic moment and can hence provide important information. Applying the sum rules to the Ir-$L_{2,3}$ XMCD spectrum of Sr$_2$Co$_{0.5}$Ir$_{0.5}$O$_4$ gives a ratio $L_z/(2S_z+7T_z) = 0.45(1)$ for Ir$^{5+}$. This value is close to the ratio $L_z/2S_z = 0.5$ predicted for a $J_{eff}=0$ system, if one neglects $T_z$. However, neglecting $T_z$ in iridates can be very misleading. In fact, taking in account that $T_z$ increases going from 3$d$ to 4$d$ and, further, 5$d$ transition metals and that $S=1$ for LS Ir$^{5+}$ ions, then $7T_z$ in Sr$_2$Co$_{0.5}$Ir$_{0.5}$O$_4$ might be actually comparable to $2S_z$.

In order to circumvent this uncertainty problem related to $T_z$, we have performed configuration -interaction cluster calculations using the Quanty Package ~\cite{Haverkort.2012,Lu.2014,Haverkort.2014}. The desired information can then be directly extracted from these calculations once the calculations can successfully produce an accurate simulation of the experimental XAS and XMCD spectral line shapes. The method uses an IrO6 cluster, which includes explicitly the full atomic multiplet interaction, the hybridization of Ir with the ligands, the crystal field acting on the Ir ion and the non-cubic crystal field acting on the ligands. The hybridization strengths and the crystal field acting on the oxygen ligands were extracted \textit{ab-initio} by DFT calculations carried out using the full-potential local-orbital code \texttt{FPLO} \cite{fplo1}. The non cubic crystal field acting on the Ir ion was fine tuned to best fit the experimental XAS and XMCD spectra. The parameters used in the calculations are listed in reference ~\cite{parameters.2017}. Since we are dealing with a polycrystalline sample, we simulated the experimental data by summing two calculated spectra: one for light with the Poynting vector in the xy plane and one with the Poynting vector along the z-axis, with a weighting ratio 2:1. As explained in a more detailed way later, an exchange field of 16~meV parallel to the magnetic field was introduced in the Hamiltonian in order to reproduce the size of the experimental XMCD signal.

The calculated Ir-$L_{2,3}$ XAS and XMCD spectra are plotted in Fig.~\ref{fig.1} as solid red and blue
curves, respectively. One can clearly see that the line shapes of the measured Ir-$L_{2,3}$ spectra are very well reproduced by our simulations. The nice agreement between theory ad experimental data is
also quantitative: the calculated isotropic \cite{isotropic.2017} ratio $L_z/(2S_z+7T_z)=0.46$ is essentially identical to the value of 0.45 extracted from the application of the sum rules to our experimental spectra.

\begin{table}[ht]
\caption{Weight of various configurations for the Ir$^{5+}$ ground state.}
\label{tab.1}
%	\centering
\begin{center}
		\begin{tabular} {c c c c c c c c}\hline\hline
%\begin{tabular}{lcr}
%			\hline\hline
          \\
%		 H$_{ex}$ (meV) & 5$d^4$ & 5$d^5\underline{L}$ & 5$d^6\underline{L}^2$ & 5$d^7\underline{L}^3$ \\	\hline
		 5$d^4$ & 5$d^5\underline{L}$ & 5$d^6\underline{L}^2$ & 5$d^7\underline{L}^3$ \\	\hline
		  7.0\% & 30.8\% & 43.3\% & 18.8\% \\
			\hline\hline
\end{tabular}
\end{center}
\end{table}

The best fit to our experimental spectra is obtained for the $t_{2g}$ orbitals split by an effective
tetragonal crystal field of $\Delta_{t2g}^{eff}=-325$~meV, where this effective crystal field includes the effect of the hybridization with the oxygen ions. The negative sign indicates that the $d_{xy}$ orbital is lower in energy than the $d_{xz}$ and $d_{yz}$ ones. A similar negative $\Delta_{t2g}^{eff}$ was observed also in the case of Sr$_2$IrO$_4$ \cite{AgrestiniRIXS.2017,Bogdanov.2015}. The value of $\Delta_{t2g}^{eff}$ is more than 10 times larger than the trigonal $t_{2g}$ splitting (10-20 meV) estimated for the double perovskite Sr$_2$YIrO$_6$ \cite{Bhowal.2015}. Such a large splitting is of the same order of the SOC ($\sim$0.4 eV) and should have consequences for the magnetic properties. Indeed, our calculations reveal that there is a strong anisotropic effect: the ratio between the orbital and spin moments is $L_x/2S_x$ = 0.35 and $L_z/2S_z$ = 0.49 for magnetic field applied in the $xy$ plane and along the $z$-axis, respectively. Furthermore, the calculated intra-atomic magnetic dipole moment was found to be large along the $z$-direction, i.e. $T_z/S_z = -0.35$, and nearly negligible in the xy plane. i.e. $T_x/S_x = -0.03$. These are the values obtained in the presence of the 16 meV exchange field. Switching off this exchange field, we obtain $L_x/2S_x$ = 0.43 and $L_z/2S_z$ = 0.78, the values more relevant for low field experiments.

%\begin{largetable}
\begin{table*}[t]
\caption{Calculated $J_{eff}$, orbital-to-spin ratio, and magnetic susceptibility for different scenarios.}
\label{tab.2}
%\begin{center}
% \begin{adjustbox}{max width=\textwidth}
% \resizebox{\columnwidth}{!}{%
\begin{tabular} {*{9}{c}}
%\begin{tabular}{lcr}
\hline\hline
		 local & 10$Dq^{eff}$ & cov. & $J_{eff}$ & $L_x/2S_x$ & $L_z/2S_z$ & $\chi_x$ & $\chi_z$  \\	
		 symm. & eV &  &  & $B // x$ & $B // z$ & (emu/mole/Oe) & (emu/mole/Oe) \\	
             \hline
		 $O_h$ & 10.0 & ionic &  0.07 & 0.49 & 0.49 & 1.0$\times 10^{-3}$ & 1.0$\times 10^{-3}$  \\
		 $O_h$ &  3.0 & ionic &  0.69 & 0.38 & 0.38 & 6.8$\times 10^{-4}$ & 6.8$\times 10^{-4}$  \\
		 $D_{4h}$ & 10.0 & ionic &  0.34 & 0.40 & 0.59 & 1.3$\times 10^{-3}$ & 4.9$\times 10^{-4}$    \\
		 $D_{4h}$ & 3.0 & ionic &  0.77 & 0.28 & 0.63 & 8.2$\times 10^{-4}$ & 4.3$\times 10^{-4}$ \\
		 $O_h$ & 10.0 & covalent &  0.06 & 0.68 & 0.68 & 3.5$\times 10^{-4}$ & 3.5$\times 10^{-4}$   \\
		 $O_h$ & 3.0 & covalent &  0.29 & 0.54 & 0.54 & 8.1$\times 10^{-4}$ & 8.1$\times 10^{-4}$    \\
		 $D_{4h}$ & 10.0 & covalent &  0.77 & 0.31 & 0.92 & 9.2$\times 10^{-4}$ & 1.7$\times 10^{-4}$    \\
		 $D_{4h}$ & 3.0 & covalent &  0.53 & 0.43 & 0.78 & 1.1$\times 10^{-3}$ & 3.9$\times 10^{-4}$   \\
		 \hline\hline
\end{tabular}%
%}
%\end{adjustbox}
%\end{center}
\end{table*}

Furthermore, also the parameters used in the calculations to obtain a good fit to the experimental
spectra reveal that the Sr$_2$Co$_{0.5}$Ir$_{0.5}$O$_4$ system is strongly covalent and thus far
from ionic. Consistent with the high-valence state of the Ir ion, the charge transfer energy is
negative, i.e. $\Delta_{CT} \sim -1.5$~eV. The consequence is that only 7.1\% of the ground state
of the Ir$^{5+}$ ion has in fact the 5$d^4$ character, while the configurations 5$d^5\underline{L}$
and 5$d^6\underline{L}^2$, where $\underline{L}$ denotes a ligand hole, are dominant. See Table 1.
On average, the number of electrons in the $d$ bands is $n_e=5.73$, i.e. almost 2 electrons are
transferred from the oxygens to the Ir ions.

The question now arises to what extent the $J_{eff}=0$ state is an accurate description of the ground state of the Ir$^{5+}$ ion in Sr$_2$Co$_{0.5}$Ir$_{0.5}$O$_4$ in view of the presence of the large tetragonal field splitting, strong covalency, and participation of the $e_g$ orbitals in addition to the $t_{2g}$. To this end it is instructive to calculate with full atomic multiplet theory the relevant expectation values of the Ir ion quantum numbers for several scenarios as listed in Table 2 (no exchange field). Starting with an ionic Ir$^{5+}$ 5$d^5$ ion in octahedral symmetry ($O_h$) with a large value for the octahedral crystal field splitting of 10$Dq$ = 10 eV, we find that $J_{eff}$ is 0.07. Here we defined $\mathbf{J_{eff}}$ = $\mathbf{L_{eff}}$ + $\mathbf{S}$, where the $\mathbf{L_{eff}}$ operator is obtained by rotating the orbital basis of the $\mathbf{L}$ operator to the cubic harmonics. The rotation matrix was modified to only keep the $t_{2g}$ subset of the $d$ eigen-orbitals. After projecting out the $e_g$ orbitals, the angular momentum operator is rotated back to the spherical harmonics. As the covalence mixes the $d$ and ligand orbitals, in order to find good quantum numbers we calculated the expectation value of the $\mathbf{J_{eff}^2}$ operator acting on both the Ir-$d$ and ligand-$d$ shell, i.e. acting on the total IrO$_6$ cluster. While the ideal $J_{eff}$ =0 is the value one obtains when only the $t_{2g}$ orbitals span the Hilbert space, i.e when the $e_g$ orbitals are completely projected out by making 10$Dq$ infinitely large, the $J_{eff}$ = 0.07 value for 10$Dq$ = 10 eV indicates that this is already close to the ideal situation. We have also calculated the $L_x/2S_x$ (and $L_z/2S_z$) ratio and found a value of 0.49, which is very close to the expected 0.50 number for the pure $J_{eff}=0$ state. The magnetic susceptibility is calculated at 1.0 x 10$^{-3}$ emu/mole/Oe.

Next we lower the octahedral crystal field splitting to the value we find in Sr$_2$Co$_{0.5}$Ir$_{0.5}$O$_4$, namely  10$Dq$ = 3 eV. See Table 2. We find $J_{eff}$ = 0.69, which indicates that we are far away from the ideal $J_{eff}=0$ state. Still being in the ionic model, i.e. the hybridization with the oxygen ligands have not been included, this finding indicates that the $e_g$ orbitals contribute significantly to the ground state of the Ir$^{5+}$ ion. This mixing in of the $e_g$ orbitals does not take place on the one-electron level since $e_g$ and $t_{2g}$ belong to different  irreproducible representations in $O_h$, but it does take place on the multi-electron level via the full atomic multiplet interactions. These multiplet interactions, characterized by the Slater $F^2$ and $F^4$ integrals, are indeed not at all small compared to the $e_g$-$t_{2g}$ crystal field splitting ($10Dq$) and their effect cannot be ignored. The $L_x/2S_x$ (and $L_z/2S_z$) ratio reduces to 0.38, and also the magnetic susceptibility decreases to 6.8 x 10$^{-4}$ emu/mole/Oe, i.e. numbers that deviate strongly from those of the pure $J_{eff}=0$ state.

The influence of the non-cubic crystal field on the $J_{eff}=0$ state is also listed in Table 2. The calculations were performed in $D_{4h}$ symmetry. Already in the ionic and large 10$Dq$ limit we can observe that the tetragonal crystal field as it is present in Sr$_2$Co$_{0.5}$Ir$_{0.5}$O$_4$ causes $J_{eff} = 0.34$ together with a strong anisotropy in the magnetic properties: $L_x/2S_x$ = 0.40 vs. $L_z/2S_z$ = 0.59, $\chi_x$ = 1.3 x 10$^{-3}$ vs. $\chi_z$ = 4.9x 10$^{-4}$ emu/mole/Oe. Still in the ionic limit but now reducing to the realistic 10$Dq$ = 3 eV value, the tetragonal crystal field brings the system truly far away from the  $J_{eff}=0$ situation: $J_{eff} = 0.77$ together also with strong anisotropy, i.e. $L_x/2S_x$ = 0.28 vs. $L_z/2S_z$ = 0.63, $\chi_x$ = 8.2 x 10$^{-4}$ vs. $\chi_z$ = 4.3x 10$^{-4}$
emu/mole/Oe.

The effect of covalency is also systematically investigated in Table 2. Keeping the same effective
octahedral and tetragonal crystal field splittings as in the ionic calculations, we can observed that
the hybridization of the Ir $5d$ orbitals with the O $2p$ ligands has a strong effect on the values
for all relevant quantum numbers of the Ir ion: $J_{eff}$, $L_x/2S_x$, $L_z/2S_z$, $\chi_x$, and
$\chi_z$ all deviate appreciably from the ionic case. It is difficult to find a trend here and the only
message we can learn from Table 2 is that one has to calculate it explicitly for each case of interest.

\begin{figure}[t]\centering
\includegraphics[width=\linewidth]{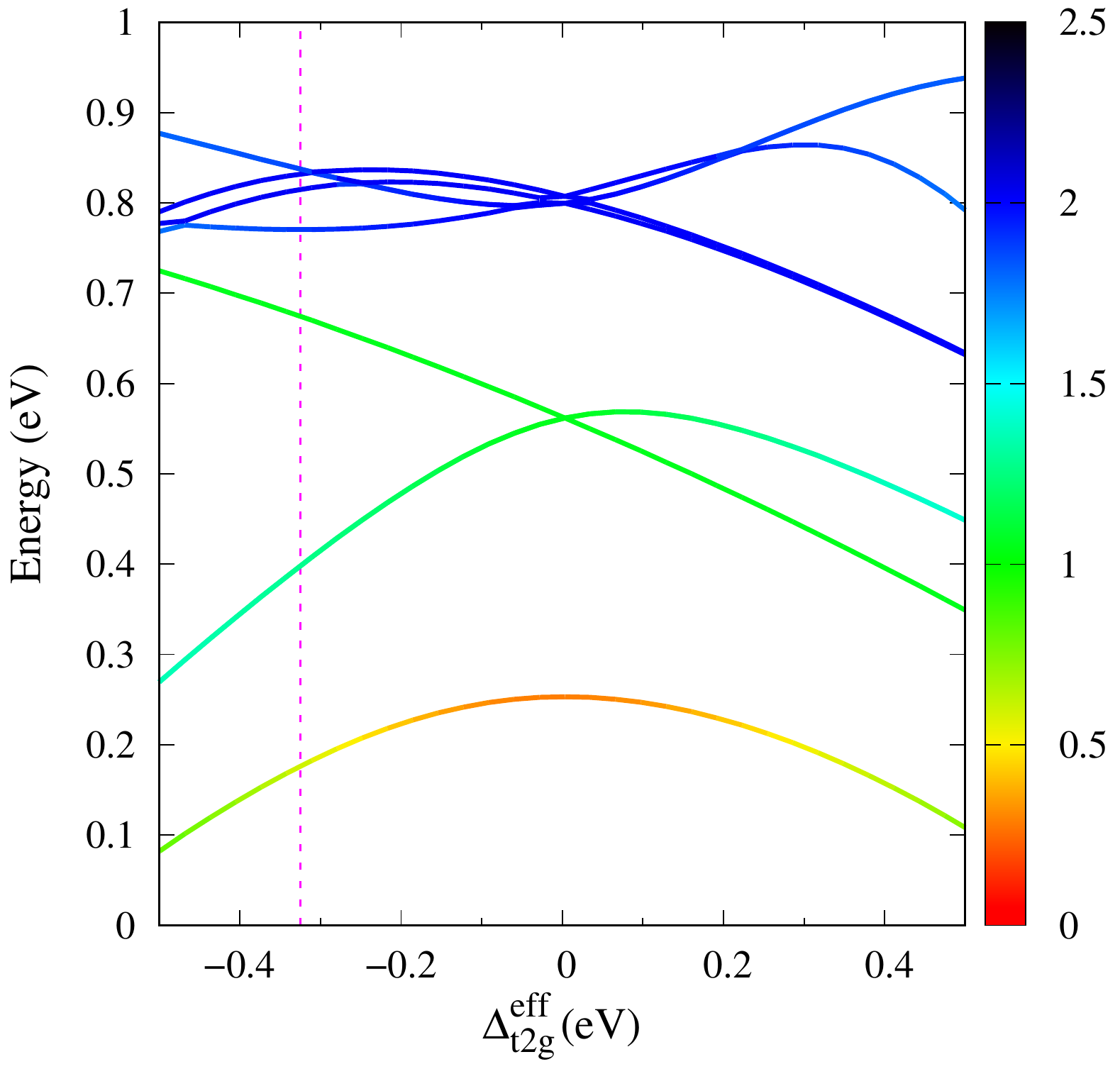}
\caption{ Energy level diagram of the Ir$^{5+}$ ($d^4$) ion as a function of the effective tetragonal
crystal field in a $D4h$ local symmetry and $10Dq^{eff}=3.0$~eV. The vertical magenta line indicates the $\Delta_{t2g}^{eff}$ of Sr$_2$Co$_{0.5}$Ir$_{0.5}$O$_4$ as obtained by the simulation of the XAS and XMCD spectra. The colours line represent the evolution of the expectation value of $J_{eff}$ versus $\Delta_{t2g}^{eff}$. The value of $J_{eff}$ ranges from 0 (red) to 2.5 (black) as indicated by the palette on the right.
}\vspace{-0.2cm}
\label{fig.2}
\end{figure}

Given the fact that the $J_{eff} = 0$ state is no longer valid in the presence of finite octahedral
crystal field splitting, strong tetrahedral crystal field interaction, as well as covalency, we now
study to what extent the Ir$^{5+}$ ion can still be described as a singlet Van Vleck paramagnetic
ion. In Fig. 2 we show the energy level diagram of the Ir$^{5+}$ ion in the IrO$_6$ cluster as
a function of the effective tetragonal crystal field using otherwise the parameters we found for
Sr$_2$Co$_{0.5}$Ir$_{0.5}$O$_4$. The tetragonal distortion causes a splitting of the triplet
$J_{eff}=1$ excited state. However, the singlet ground state is still well below the first excited state $J_{eff}=1$ even for very large values of $\Delta_{t2g}^{eff}$. So for Sr$_2$Co$_{0.5}$Ir$_{0.5}$O$_4$ where
we found that $\Delta_{t2g}^{eff}$ is of the same order as the spin-orbit coupling, the excitation gap
(235 meV) between the singlet and the triplet states is only slightly reduced from the value expected in a
cubic symmetry (3/4 SOC = 300 meV). The colors of the curves in Fig. 2 indicate the value of
$J_{eff}$ as described in the above sections.

\begin{figure}[t]\centering
\includegraphics[width=\linewidth]{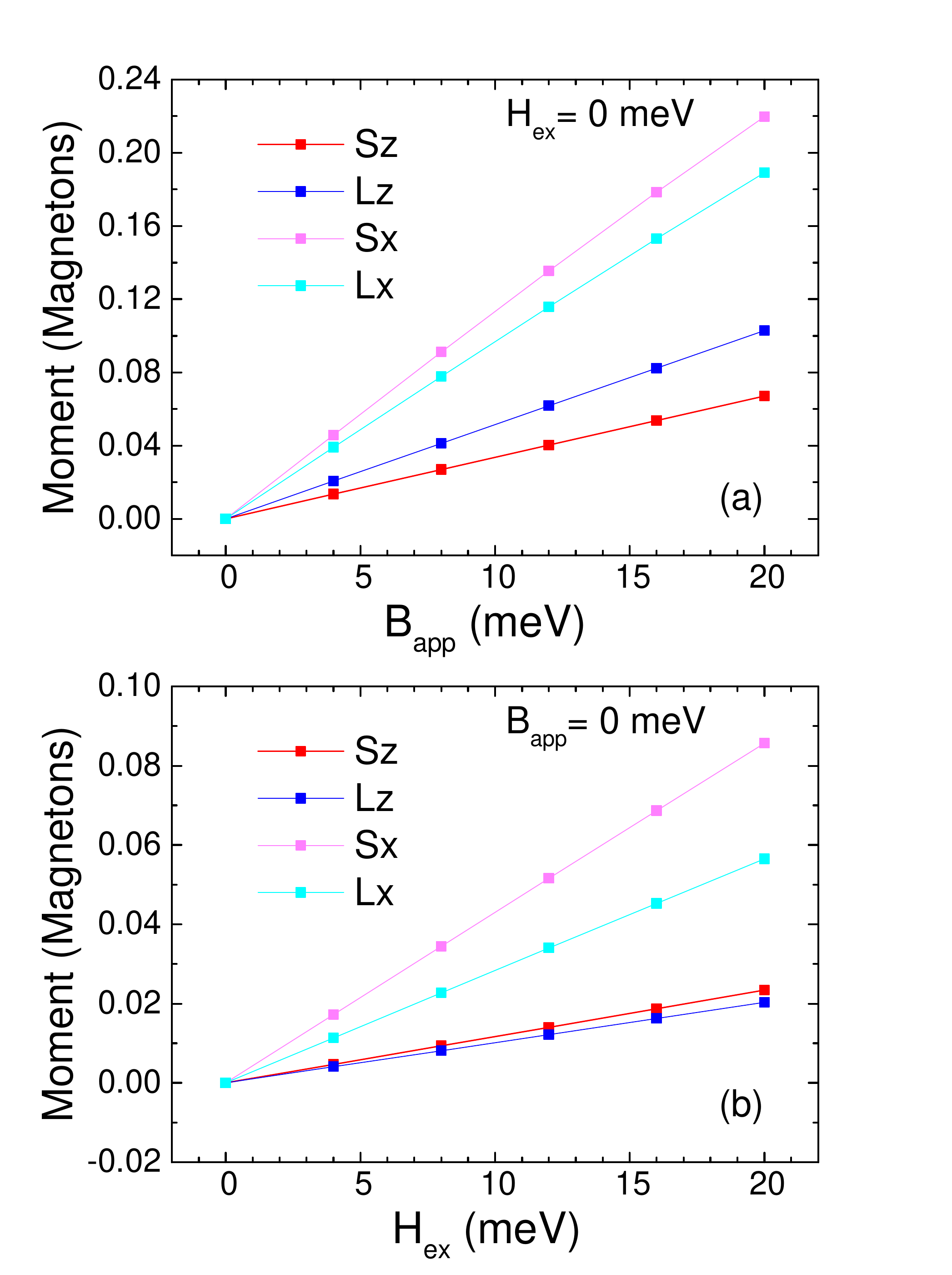}
\caption{Calculated spin and orbital moments as a function of the applied field (a) or exchange field (b).
}\vspace{-0.2cm}
\label{fig.3}
\end{figure}

Next we calculate the magnetic properties of the Ir$^{5+}$ ion versus magnetic field. As shown in Fig. 3, the calculated spin and orbital moments (and the XMCD signal) are zero for both the applied magnetic field and the exchange field equal to zero, and increase linearly as a function of the applied field or exchange field. This is consistent for a Van Vleck system. We have also calculated the susceptibility as a function of temperature. In the calculations we considered the thermal population of the Ir$^{5+}$ energy levels using a Boltzmann distribution. The susceptibility was found to be temperature independent with a value of $\chi=8.3 \times 10^{-4}$ emu/mole/Oe, which is the isotropic average of the values listed in Table 2. This calculated value is in nice agreement with the value of the Van Vleck susceptibility measured by SQUID
in Ba$_2$YIrO$_6$ ($\chi_{VV} =7.51 \times 10^{-4}$ emu/mol/Oe)\cite{Dey.2016},
Sr$_2$YIrO$_6$ ($\chi_{VV} =6.6 \times 10^{-4}$ emu/mol/Oe) \cite{Corredor.2017},
and NaIrO$_3$ ($\chi_{VV} =19 \times 10^{-4}$ emu/mol/Oe) \cite{Bremholm.2011}.
Hence, from the above results we conclude that, despite the $J_{eff}=0$ state is quite perturbed, the
Ir$^{5+}$ ions in Sr$_2$Co$_{0.5}$Ir$_{0.5}$O$_4$ are still describable as Van Vleck ions.

\begin{figure}[t]\centering
\includegraphics[width=\linewidth]{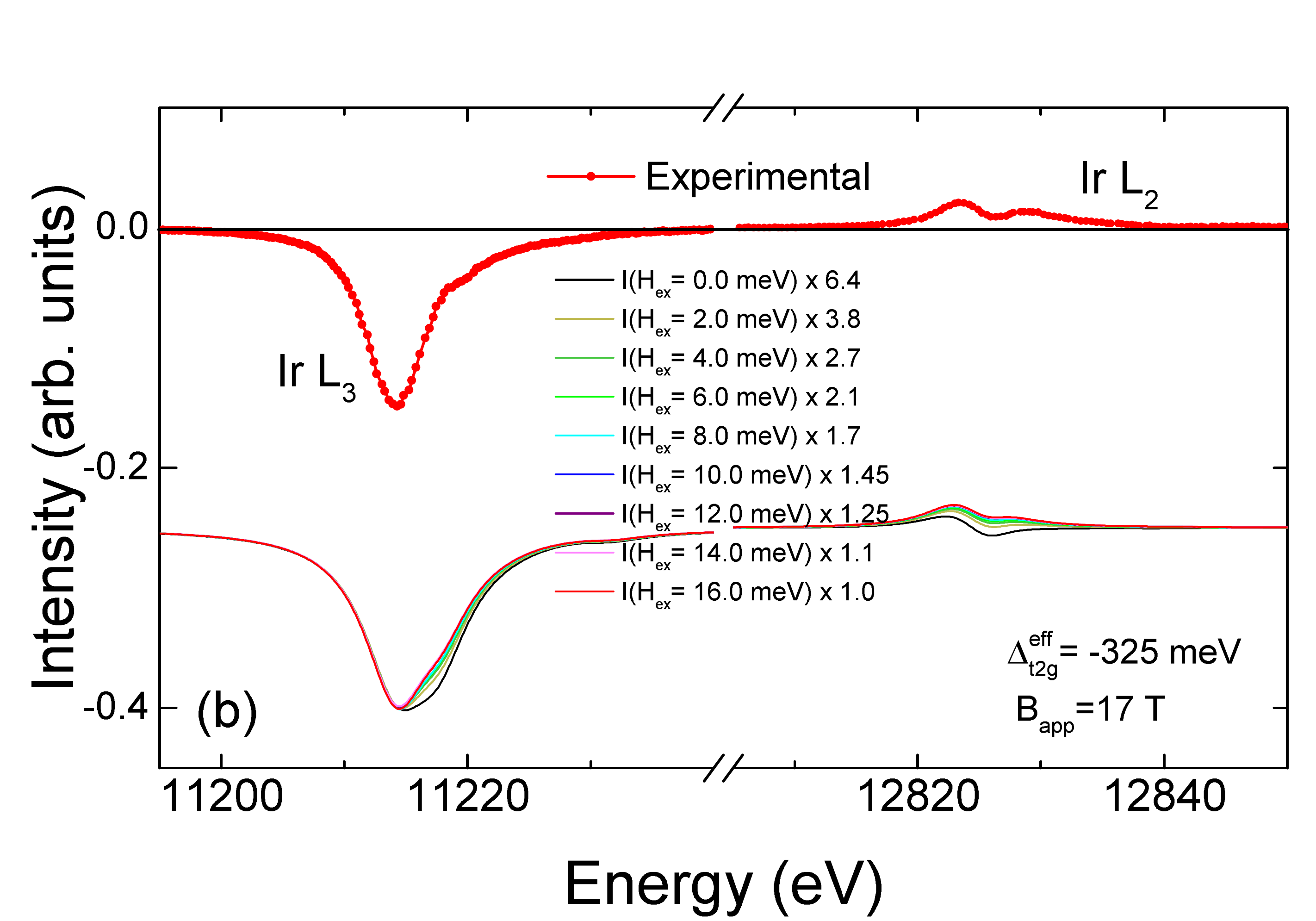}
\caption{ Ir-$L_{2,3}$ XMCD simulations (solid lines) calculated using different values of $H_{ex}$, together with the experimental XMCD spectrum of Sr$_2$Co$_{0.5}$Ir$_{0.5}$O$_4$ (red circles). For a better comparison of the line shapes, the simulated spectra are normalized to the XMCD peak height, with the normalization factor also indicated in the legend.}\vspace{-0.2cm}
\label{fig.4}
\end{figure}

Finally, we would like to discuss about the exchange field in our simulations. We introduced in the
Hamiltonian an exchange field parallel to the magnetic field, since the calculated XMCD signal in an
applied field of 17 Tesla is 6 times smaller than the measured one. An exchange field of about 16~meV
is needed in order to reproduce the size of the experimental XMCD spectrum. This is illustrated
in Fig. 4, where we show the XMCD spectra calculated for 17 T magnetic field plus the presence of
varying strengths of the exchange field. It is also important to note that the introduction of a strong exchange field is necessary to obtain a good fit of to the experimental XMCD spectrum: for
zero exchange field the line shape of the XMCD spectrum cannot be properly simulated, both at the
$L_3$ and $L_2$ edges, no matter the value of the tetragonal distortion.

\begin{figure}[t]\centering
\includegraphics[width=\linewidth]{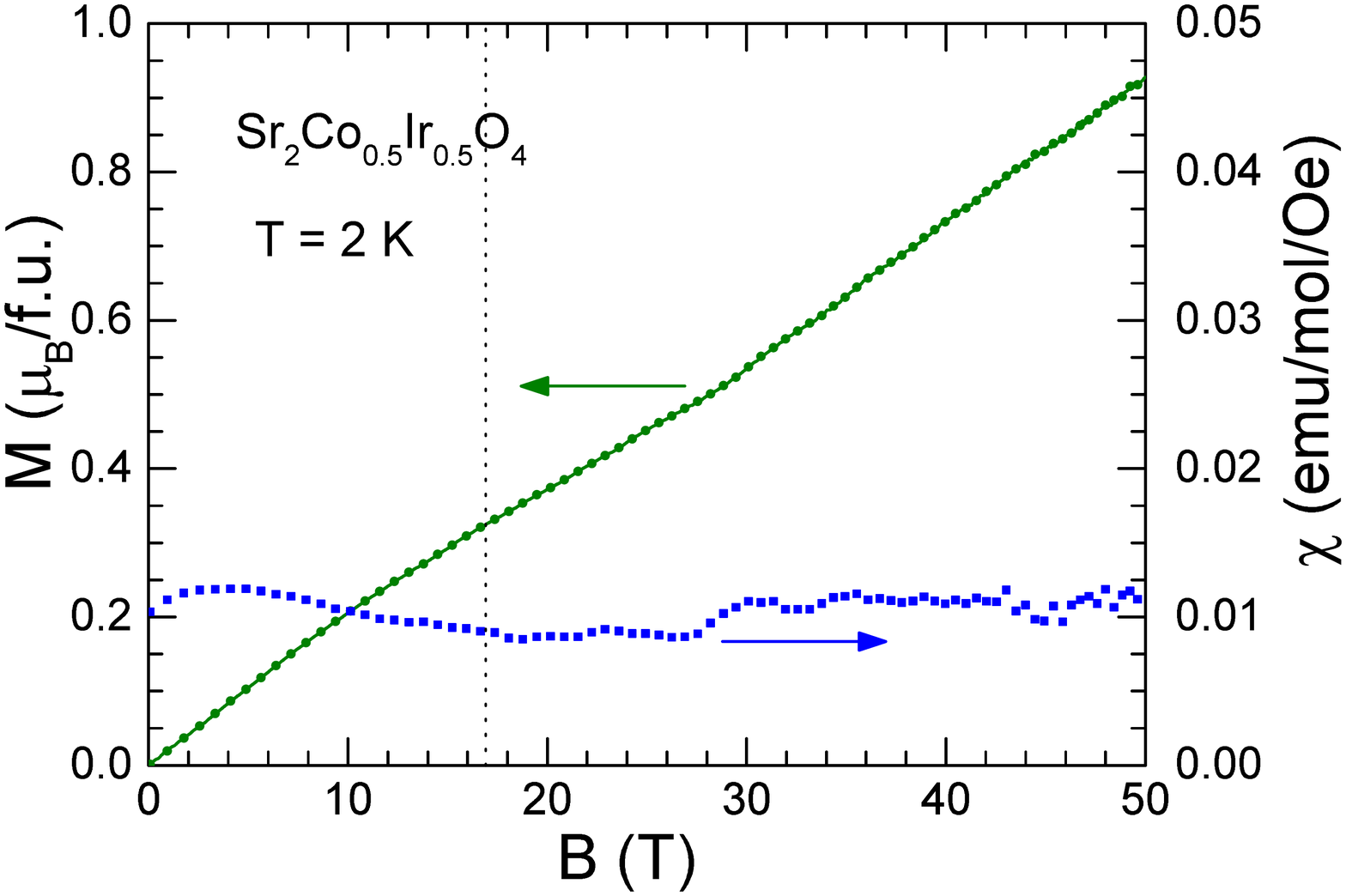}
\caption{Magnetization of Sr$_2$Co$_{0.5}$Ir$_{0.5}$O$_4$ as a function of magnetic field,
together with the derived magnetic susceptibility.}\vspace{-0.2cm}
\label{fig.5}
\end{figure}

To unveil the origin of the 16 meV exchange field in our 17 T XMCD experiment, we measured the
magnetization and magnetic susceptibility of our Sr$_2$Co$_{0.5}$Ir$_{0.5}$O$_4$ sample using pulsed magnetic fields up to 58T at 2K. The results are displayed in Fig. 5. We can see a practically regular linear
increase of the magnetization with field, yielding about 0.3 $\mu_B$ per formula unit (f.u.) at 17 T.
The magnetic susceptibility varies around 0.01 emu/mole/Oe over the entire magnetic field range.
This value is an order of magnitude larger than the magnetic susceptibility of the Ir$^{5+}$ ion,
which was calculated to be $8.3 \times 10^{-4}$ emu/mole/Oe on the basis of our XMCD analysis,
with similar values from the Ba$_2$YIrO$_6$ \cite{Dey.2016}, Sr$_2$YIrO$_6$ \cite{Corredor.2017},
and NaIrO$_3$ \cite{Bremholm.2011} compounds. This in turn implies that the large magnetic
susceptibility of the Sr$_2$Co$_{0.5}$Ir$_{0.5}$O$_4$ compound is caused by mostly the high-spin
Co$^{3+}$ ions, and that the 0.3 $\mu_B$ per f.u. magnetization at 17 T is associated with the
canting of these antiferromagnetically ordered Co ions.

With 0.5 Co per f.u. we thus find that the 17 T magnetic field induces a canted magnetic moment
of about 0.6 $\mu_B$ per Co. The presence of 16 meV exchange field at 17 T felt by the Ir ions
can then be attributed to the presence of canted moments at the Co sites. With each Ir ion
coordinated by four nearest-neighbor Co ions, we then have an exchange field of 4 meV per Co
neighbor having 0.6 $\mu_B$ moment, i.e. about 7 meV per neighbor$\cdot\mu_B$. This seems to be not so unreasonable when considering, for example, the case of NiO, where an exchange field is found of 19 meV per Ni neighbor with
2 $\mu_B$ \cite{Dietz.1971,Hutchings.1972}, i.e. about 9.5 meV per neighbor$\cdot\mu_B$. The considerations we
just have made should of course be made self-consistent. With the 16 meV exchange field, we calculate
that the Ir ions acquire about 0.16 $\mu_B$ magnetic moment, i.e. 0.08 $\mu_B$ per f.u. This leaves 0.3 $\mu_B$ - 0.08 $\mu_B$ = 0.22 $\mu_B$ moment for the Co ions per f.u., or 0.44  $\mu_B$ per Co ion, i.e. about 9 meV per neighbor$\cdot\mu_B$. The exchange field strength per neighbor per $\mu_B$ is then quite similar to the NiO case. However, the agreement
is likely to be fortuitous considering the fact that we have not evaluated the energetics of the
virtual excitations involved in these super exchange type of interactions. Nevertheless, the numbers
are not unreasonable and may serve as a first order estimate.

\section{Conclusions}
We have investigated the local magnetism of the Ir$^{5+}$ ions in the layered hybrid solid state oxide
Sr$_2$Co$_{0.5}$Ir$_{0.5}$O$_4$ by employing a combined experimental and theoretical x-ray magnetic
circular dichroism (XMCD) spectroscopy study at the Ir-$L_{2,3}$ edges. From simulations of the experimental XMCD spectrum we found that the orbital-to-spin moment
ratio is significantly reduced compared to the value expected for a pure $J_{eff}=0$ ground state.
We show that the combination of atomic multiplet interactions, large tetragonal crystal field, and high
covalency brings the system away from the ideal $J_{eff}=0$ scenario. Nevertheless our calculations show
also that the excitation gap between the singlet ground state and the triplet excited state is still very large, and
that the Ir$^{5+}$ ions exhibit magnetic properties, as a function of both temperature and applied field,
which are typical for Van Vleck ions.

\begin{acknowledgments}
We gratefully acknowledge the ESRF staff for providing beamtime. The research in Dresden was partially supported by the Deutsche Forschungsgemeinschaft (DFG) through SFB 1143 and by the Bundesministerium f\"ur Bildung und Forschung (BMBF), project grant number 03SF0477B (DESIREE). K. Chen benefited from support of the DFG via Project SE 1441. We acknowledge the support of the Hochfeld-Magnetlabor Dresden (HLD) at the Helmholtz-Zentrum Dresden-Rossendorf (HZDR), member of the European Magnetic Field Laboratory (EMFL).
\end{acknowledgments}

\bibliography{Sr2IrCoO4_PRB_20180217}

\begin{thebibliography}{38}
\bibitem[{par()}]{parameters.2017}
\bibinfo{note}{{IrO6 cluster parameters [eV]: $U_{dd}=1.0$, $U_{pd}=2.0$, ionic
  crystal field 10$D_q=1.6$, $\Delta t_{2g}= -0.27$, $\Delta e_g$= -0.35,
  charge transfer energy $\Delta_{CT}= -1.5$, hybridization $V(b_{1g}) = 5.30$,
  $V(a_{1g}) = 4.93$, $V(b_{2g}) = 2.70$, $V(e_g) = 2.64$, ligand-crystal field
  10$Dq^{lig} = 0.43$, $\Delta_{eg}^{lig} = -1.10$, $\Delta_{t2g}^{lig} =
  -0.54$, spin-orbit coupling $ \zeta_{5d}=0.4$, exchange field $H_{ex} =
  0.016$ and magnetic field 17 T. The Slater integrals were reduced to 70\% of
  Hartree-Fock values. The effective splittings of the $5d$ orbitals were
  estimated to be 10$Dq^{eff}=3.0$, $\Delta t_{2g}^{eff}=-0.325$, $\Delta
  e_g^{eff}=-0.265$}}.
\bibitem[{iso()}]{isotropic.2017}
\bibinfo{note}{{Taking in account that the sample is polycrystalline, the
  isotropic value of the $(L_z/(2S_z+7T_z)$ ratio was calculated as
  $(L_z+2L_x)/(2(S_z+2S_x)+7(T_z+2T_x))$}}.
\end{thebibliography}

\end{document}